# Scaling Artificial Intelligence for Prostate Cancer Detection on MRI towards Organized Screening and Primary Diagnosis in a Global, Multiethnic Population (Study Protocol)

Anindo Saha, Joeran S. Bosma, Jasper J. Twilt, Alexander B.C.D. Ng, Aqua Asif, Kirti Magudia, Peder Larson, Qinglin Xie, Xiaodong Zhang, Chi Pham Minh, Samuel N. Gitau, Ivo G. Schoots, Martijn F. Boomsma, Renato Cuocolo, Nikolaos Papanikolaou, Daniele Regge, Derya Yakar, Mattijs Elschot, Jeroen Veltman, Baris Turkbey, Nancy A. Obuchowski, Jurgen J. Fütterer, Anwar R. Padhani, Hashim U. Ahmed, Tobias Nordström, Martin Eklund, Veeru Kasivisvanathan, Maarten de Rooij, Henkjan Huisman, on behalf of the PI-CAI–ProCAncer-I–COMFORT–STHLM3-MRI–PRIME consortia*

## Study Overview

### Background

Prostate cancer is a genomically diverse disease with a broad spectrum of outcomes. One million receive a diagnosis and 300,000 die from clinically significant prostate cancer each year worldwide —with a 2024 Lancet Commission projecting a doubling in the number of annual cases over the next 20 years.[1,2] MRI has an increasingly important role in the early diagnosis of prostate cancer and has been recommended prior to biopsies by clinical guidelines in Europe, UK and the USA.[3-5] Radiologists follow the Prostate Imaging–Reporting and Data System version 2.1 (PI-RADS v2.1) —a standardized, semi-quantitative assessment to read prostate MRI that mandates substantial expertise for proper usage.[6] Although PI-RADS v2.1 has proven effective in reducing biopsies, it remains susceptible to low inter-reader agreement, sub-optimal interpretation and overdiagnosis.[7-9] At the same time, a global shortage of experienced radiologists persists and healthcare systems around the world are struggling to meet the rising demand in imaging.[10-13] Addressing this shortage is crucial for sustaining equitable care in routine practice and facilitating population-based prostate cancer screening programs through the future.[14,15]

Over the past decade, modern artificial intelligence (AI) systems have matched expert clinicians in medical image analysis across several specialties including dermatology, ophthalmology, histopathology and radiology.[16-19] In the July 2024 issue of The Lancet Oncology, the PI-CAI consortium published the primary outcomes of the largest diagnostic accuracy study for prostate cancer detection on MRI reported till date.[20] Using a retrospective cohort of 10,207 prostate MRI examinations from four European tertiary care centers, we presented level 2b evidence that a state-of-the-art AI system was statistically superior in clinically significant prostate cancer detection in comparison to a pool of 62 radiologists using PI-RADS v2.1 (45 centers, 20 countries; median 7 [IQR 5–10] years of experience in reading prostate MRI). However, whether the AI system would demonstrate similar diagnostic performance in other settings (e.g., in a population-based screening setting, in clinics with low imaging quality or a non-standardized imaging protocol, in a multiethnic population) was not confirmed, thereby limiting its impact from potentially being practice-changing.

### Primary Endpoints and Secondary Measures

In this intercontinental, confirmatory study, we include a retrospective cohort of 22,481 MRI examinations (21,288 patients; 46 cities in 22 countries) to train and externally validate the PI-CAI-2B model, i.e., an efficient, next-generation iteration of the state-of-the-art AI system that was developed for detecting Gleason grade group ≥2 prostate cancer on MRI during the PI-CAI study.[20] Of these examinations, 20,471 cases (19,278 patients; 26 cities in 14 countries) from two EU Horizon projects (ProCAncer-I, COMFORT) and 12 independent centers based in Europe, North America, Asia and Africa, are used for training and internal testing.[20-22] Additionally, 2010 cases (2010 patients; 20 external cities in 12 countries) from population-based screening (STHLM3-MRI, IP1-PROSTAGRAM trials) and primary diagnostic settings (PRIME trial) based in Europe, North and South Americas, Asia and Australia, are used for external testing.[23-25] Primary endpoint is the proportion of AI-based assessments in agreement with the standard of care diagnoses (i.e., clinical assessments made by expert uropathologists on histopathology, if available, or at least two expert urogenital radiologists in consensus; with access to patient history and peer consultation) in the detection of Gleason grade group ≥2 prostate cancer within the external testing cohorts. Our statistical analysis plan is prespecified with a hypothesis of diagnostic interchangeability to the standard of care at the PI-RADS ≥3 (primary diagnosis) or ≥4 (screening) cut-off, considering an absolute margin of 0.05 and reader estimates derived from the PI-CAI observer study (62 radiologists reading 400 cases). Secondary measures comprise the area under the receiver operating characteristic curve (AUROC) of the AI system stratified by imaging quality, patient age and patient ethnicity to identify underlying biases (if any).

We advise caution when interpreting the outcomes of the primary comparison in this study, i.e., a statistical evaluation of whether the assessments of the PI-CAI-2B AI system are diagnostically interchangeable to the standard of care within the external testing cohorts —rather than an evaluation of whether its assessments are diagnostically non-inferior to the local radiology readings that were historically made for these cases during routine practice. We opted to forgo the latter analysis as the primary comparison (as conventionally practiced and done before during the PI-CAI study) due to the following considerations:

- A valid reference standard is not available for all cases within the external testing cohorts, i.e., there is no histopathology evidence or follow-up outcomes for the vast majority of negative MRI examinations (as per standard diagnostic practice). Thus, to avoid partial reference standard bias, we chose not to use diagnostic accuracy as our primary endpoint but rather to use an interchangeability metric. Please note that interchangeability testing allows the inclusion of all cases, whereas diagnostic accuracy testing would necessitate the exclusion of cases without a valid reference standard.

- Demonstrating interchangeability with respect to the standard of care diagnoses at external testing sites is critical for large-scale clinical implementation —for which, comparing the assessments of an AI system to that of the standard of care following the interpretation of an average radiologist becomes more relevant, rather than comparing them to the consensus observations of two specific radiologists from a clinical trial.

When calculating the estimates for an average radiologist, we take several measures to control for the effects of inter-reader agreement and the prevalence of Gleason grade group ≥2 cancer in a given testing cohort. For transparency, we also estimate the diagnostic accuracy of the AI system in comparison to that of the historical radiology readings within the external testing cohorts as secondary measures, using multiple imputation to adjust for verification bias. For more details, please refer to the statistical analysis plan provided later in this protocol.

## Study Population

### Patient Inclusion Criteria

Inclusion criteria [23-25]:
- Adult males (≥18 years of age) suspected of harboring prostate cancer, with elevated levels of prostate-specific antigen (PSA) (≥3 ng/mL), a Stockholm3 score of ≥0.11 or abnormal findings on digital rectal exam, who subsequently underwent prostate MRI.

Exclusion criteria (training and internal testing cohorts):
- Patients who opted out of participation or did not give permission to reuse clinical data.
- Patients with non-compliant imaging (i.e., prostate gland partially visible within the field-of-view, missing or mismatched biparametric imaging sequences) or incomplete clinical outcomes (e.g., patient age, PSA level, PI-RADS scores, Gleason scores in the presence of PI-RADS ≥3 findings).

Exclusion criteria (external testing cohorts)[23-25]:
- Patients who opted out of participation or did not give permission to reuse clinical data.
- Patients with a history of prior prostate cancer (Gleason grade group ≥1) findings.
- Patients with non-compliant imaging (i.e., prostate gland partially visible within the field-of-view, missing or mismatched biparametric imaging sequences) or incomplete clinical outcomes (e.g., patient age, PSA level, PI-RADS scores, Gleason scores in the presence of PI-RADS ≥3 findings).

---

*A list of all mainline co-authors and their affiliations appears at the end of the protocol.
A list of all consortia members and their affiliations will be stated in a future publication reporting the outcomes of this study.

## Patient Cohorts

- **Primary Training and Internal Testing Cohort:** A retrospective, standardized cohort of 10,207 MRI examinations (9129 patients) that was acquired between January 1, 2012 to December 31, 2021 across four European tertiary care centers based in the Netherlands and Norway, using one of nine 1.5 or 3 Tesla scanner types from two commercial vendors (Siemens Healthineers, Philips Medical Systems). This dataset was originally acquired during the PI-CAI study, which investigated the non-inferiority of AI systems to radiologists in detecting Gleason grade group ≥2 prostate cancer detection on MRI.[20] Ethnicity data was not recorded or considered for diagnostic decision-making during routine practice at the participating centers, and thus, this information could not be retrospectively collected.

    This dataset was originally used to train (9107 cases), tune (100 cases) and test (1000 cases) the PI-CAI-1 AI system during the PI-CAI study.[20] In this study, the training set (9107 cases) will also be used to train the PI-CAI-2B system.

- **Secondary Training and Internal Testing Cohort:** A retrospective, multiethnic, intercontinental cohort of 10,264 MRI examinations (10,149 patients) that was acquired between November 29, 2012 to October 31, 2024 across 22 cities (13 countries) based in Europe, Asia, North America and Africa, using one of thirty-four 1.5 or 3 Tesla scanner types from six commercial vendors (Siemens Healthineers, Philips Medical Systems, GE HealthCare, Canon Medical Systems, AGFA HealthCare, United Imaging). This dataset was collected through two European Union: Horizon projects (ProCAncer-I: grant 952159; COMFORT: grant 101079894) and eight independent centers (Netherlands Cancer Institute, Amsterdam; Third Affiliated Hospital, Guangzhou; 108 Military Central Hospital, Hanoi; Aga Khan University Hospital, Nairobi; University of California, San Francisco Medical Center, San Francisco; University of California, Los Angeles Medical Center, Los Angeles; Isala Hospital, Zwolle; Northwestern University, Evanston).[21,22] Ethnicity data has been recorded and retrospectively classified according to a simplified representation of the OMB Directive 15 for the 2435 examinations from University of California, San Francisco Medical Center and Northwestern University.[26]

    A subset of 9264 cases from this secondary cohort, in addition to the training set of 9107 cases from the PI-CAI study (i.e., a combined training set of 18,371 cases), will be used to train the PI-CAI-2B system. The remaining held-out set of 1000 cases will constitute an internal testing cohort that will be used to validate the PI-CAI-1 and PI-CAI-2B systems in an interim analysis on data drift.

- **External Testing Cohort (PRIME Trial – Primary Diagnostic Setting):** A retrospective, multiethnic, intercontinental cohort of 476 MRI examinations (476 consecutive patients) that was acquired between March 22, 2023 to May 21, 2024 across 18 cities (11 countries) based in Europe, North and South Americas, Australia and Asia, using one of fourteen 1.5 or 3 Tesla scanner types from three commercial vendors (Siemens Healthineers, Philips Medical Systems, GE HealthCare). This dataset was originally acquired during the PRIME trial, which investigated the non-inferiority of biparametric MRI to multiparametric MRI in detecting Gleason grade group ≥2 cancer.[23] Ethnicity data has been recorded and retrospectively classified according to a simplified representation of the OMB Directive 15 classification system for this complete cohort.[26]

    This external testing cohort will be used to test one of our primary hypotheses, i.e., the PI-CAI-2B system, a state-of-the-art AI system for Gleason grade group ≥ 2 cancer detection on MRI that is adequately trained using a comprehensive dataset, will demonstrate diagnostic interchangeability to radiology readings in primary diagnostic settings across several external sites worldwide.

- **External Testing Cohort (STHLM3-MRI Trial – Population Screening):** A retrospective, standardized cohort of 1143 MRI examinations (1143 consecutive patients) that was acquired between February 5, 2018, to March 4, 2020, in Stockholm, Sweden, using one of two 1.5 or 3 Tesla scanner types from two commercial vendors (Siemens Healthineers, GE HealthCare). This dataset was acquired during the STHLM3-MRI trial, which investigated prostate cancer screening using a diagnostic strategy of blood-based risk prediction combined with MRI-targeted biopsies, in comparison to the traditional approach of using PSA testing combined with systematic biopsies.[24] Ethnicity data was not recorded or considered for diagnostic decision-making during this trial, and thus, this information could not be retrospectively collected.

    This external testing cohort will be used to test one of our primary hypotheses, i.e., the PI-CAI-2B system, a state-of-the-art AI system for Gleason grade group ≥2 cancer detection on MRI that is adequately trained using a comprehensive dataset, will demonstrate diagnostic interchangeability to radiology readings in population-based screening (where MRI is used as a second-stage tool) in an external healthcare system.

- **External Testing Cohort (IP1-PROSTAGRAM Trial – Population Screening):** A retrospective, standardized cohort of 391 MRI examinations (391 consecutive patients) that was acquired between October 10, 2018, to May 15, 2019, in London and Middlesex, United Kingdom, using one of two 1.5 or 3 Tesla scanner types from one commercial vendor (Siemens Healthineers). This dataset was acquired during the IP1-PROSTAGRAM trial, which investigated prostate cancer screening using PSA testing, MRI, and ultrasonography.[24] Ethnicity data has been recorded and retrospectively classified according to a simplified representation of the OMB Directive 15 classification system for this complete cohort.[26]

    This external testing cohort will be used to test one of our primary hypotheses, i.e., the PI-CAI-2B system, a state-of-the-art AI system for Gleason grade group ≥2 cancer detection on MRI that is adequately trained using a comprehensive dataset, will demonstrate diagnostic interchangeability to radiology readings in population-based screening (where MRI is used as a first-stage tool) in an external healthcare system.

## Procedures and Reference Standard

MRI examinations were evaluated by urogenital radiologists in compliance with PI-RADS (v1.0, v2.0 or v2.1), where each lesion was given a PI-RADS score between 1 and 5 to stratify the risk of prostate cancer (with higher scores indicating higher suspicion for clinically significant cancer). Assessments were made with access to peer consultation and multidisciplinary team analysis. Patients with positive MRI findings (i.e., in whom an area with a PI-RADS score of 3 or greater was identified) underwent biopsies. In the absence of abnormal areas on MRI (i.e., a negative result with a maximum PI-RADS score of 1 or 2), patients were not offered a biopsy or underwent systematic biopsies exclusively. Biopsies were performed by urologists, radiologists, or trained medical students, researchers or technicians under the supervision of urologists or radiologists (depending on local practice). Two to four cores were obtained for each MRI-targeted lesion. Medial or lateral cores were obtained from each sextant of the prostate gland (six to 16 cores, in total) during systematic biopsies. Hematoxylin and eosin-stained biopsy specimens were graded by uropathologists using whole-slide imaging or microscopic examination. Readings were reported using Gleason scores in compliance with the International Society of Urological Pathology guidelines.[27] Immunohistochemistry staining, patient history and peer consultation were available to aid diagnosis in adherence with the standard of care in routine practice. Deviations from this protocol, if any, took place due to scientific interventions, patient-specific factors, or changing clinical guidelines.

Within the scope of this study, clinically insignificant cancer has been defined as Gleason grade group 1 (Gleason score 6; low risk) and clinically significant cancer has been defined as Gleason grade group 2–5 (Gleason score 7–10; intermediate to very high risk). If the grade group of a lesion is discordant between findings from two different biopsy methods, then the higher grade has been applied. In the case of patients who underwent prostatectomy, grade groups determined from whole-mount specimens have been applied. Patient outcomes have been retrospectively tracked using institutional electronic health records and national registries. Patients in the external testing cohorts who were deemed to have non-elevated risk (or an absence of significant cancer) based on negative MRI findings and clinical variables, with or without histological evidence, had their diagnoses supported by a consensus of at least two expert urogenital radiologists.

## Development of the AI System

The PI-CAI-1 AI system is an equally-weighted ensemble of the top five highest performing deep learning algorithms from the PI-CAI study. The AI system was trained using 9107 cases (8029 patients; three centers in the Netherlands) and developed in parts by teams primarily based at the University of Sydney (Sydney, Australia), University of Science and Technology (Hefei, China), Guerbet Research (Villepinte, France), Istanbul Technical University (Istanbul, Türkiye), and Stanford University (Stanford, United States). Each of the five member models in the PI-CAI-1 system are ensembles of convolutional neural networks or vision transformers themselves. As such, inference using the PI-CAI-1 system requires an average of 30.5 minutes per case on typical cloud infrastructure (NVIDIA T4 GPU), and training consumes thousands of GPU hours. For a given biparametric MRI examination, the PI-CAI-1 system performs two tasks: autonomously localise and classify each lesion with clinically significant cancer (if any) using a 0–100 likelihood score and classify the overall case using a 0–100 likelihood score for clinically significant cancer diagnosis. To this end, the AI system uses imaging data and clinical metadata associated with the examination (i.e., patient age, PSA level, prostate volume) to inform its predictions.

The PI-CAI-2B AI system is an efficient, next-generation iteration of the PI-CAI-1 system. It comprises a single variant of the top-performing deep learning algorithm from the PI-CAI study (developed by Guerbet Research; Villepinte, France), that has undergone several updates during the SCALE study to improve efficiency, robustness and compatibility with larger and more heterogeneous datasets.[28] As a result, PI-CAI-2B requires an estimated 4–7 minutes for inference per case (including preprocessing, intra-sequence registration and prostate segmentation) and 800 GPU hours for training —an order of magnitude less than the PI-CAI-1 system. Additionally, PI-CAI-2B leverages a larger, scalable feature encoder to capture the rich MRI data present in its expanded training dataset of 18,371 cases (17,178 patients; 26 cities in 14 countries).

## Statistical Analysis Plan

As the primary outcome of this study, we statistically evaluate the agreement of the PI-CAI-2B AI system in Gleason grade group 2 or greater prostate cancer diagnosis on MRI, to the standard of care diagnoses within the external testing cohorts (comprising MRI examinations from the landmark PRIME, STHLM3-MRI and IP1-PROSTAGRAM trials). Here, the standard of care diagnoses represent the clinical assessments made by expert uropathologists on histopathology, if available, or at least two expert urogenital radiologists in consensus on MRI; with access to patient history and peer consultation. Statistical tests are conducted separately for each of the three trial cohorts constituting the external testing cohorts: PRIME trial cohort (primary diagnostic setting), STHLM3-MRI trial cohort (population-based screening), IP1-PROSTAGRAM trial cohort (population-based screening); to ensure that any conclusions made are specific to each of these cohort's unique characteristics and standards. Multiplicity is corrected for using the Holm–Bonferroni method.

Our statistical approach is motivated by the condition that during routine clinical practice, patients with negative MRI examinations (i.e., those assigned a maximum PI-RADS score of 2) do not generally undergo verification biopsies. As such, for approximately 33% of the patients encountered in the PRIME trial, 67% of the patients in the STHLM3-MRI trial, and 59% of the patients in the IP1-PROSTAGRAM trial, there is no gold standard reference (e.g., histopathology, follow-up outcomes) available to flag radiology assessments as false negatives when there are missed cancers, or retrospectively verify positive AI-based assessments as true positives. Therefore, we forgo the conventional practice of testing the statistical superiority in diagnostic accuracy of AI-based assessments over radiology readings in a given testing cohort, considering the partial reference and substantial verification bias within these cohorts. Instead, we investigate whether AI-based assessments within these external testing cohorts are diagnostically interchangeable to radiology readings from routine practice. We conclude interchangeability, if the AI system agrees with the standard of care diagnoses at least as often as a randomly chosen radiologist would agree with the standard of care diagnoses. For a holistic view of the evidence and its clinical context, we present these statistical outcomes alongside estimates for the inter-reader agreement of radiologists within these cohorts.[29,30]

## Primary Endpoints

### Estimation of Inter-Reader Agreement

We estimate the inter-reader agreement of radiologists in diagnosing clinically significant prostate cancer on multiparametric MRI, using the outcomes of the PI-CAI multi-reader multi-case (MRMC) observer study.[20] The PI-CAI MRMC study comprised 6174 readings from 62 radiologists (45 centers in 20 countries) across a retrospective cohort of 400 multiparametric MRI examinations from four tertiary care centers based in the Netherlands and Norway. Of these examinations, 133 cases had histologically confirmed Gleason grade group 2 or greater prostate cancer, i.e., an estimated prevalence of 33.3%. We consider two cut-offs to threshold all radiology assessments into binary decisions of 0 or 1:

- **PI-RADS ≥3**, i.e., the cut-off considered in the PI-CAI study as per standard criteria for primary diagnosis
- **PI-RADS ≥4**, i.e., the cut-off recommended for population-based screening [31]

For a case $i$, let $n_i$ be the number of readers who scored the case. Therefore, the total number of unique reader pairs can be calculated as:

$$N_i = \binom{n_i}{2} = \frac{n_i(n_i-1)}{2} \quad \text{—Eq. 1}$$

Let $A_i$ be the number of reader pairs in agreement for case $i$, where both readers made the same decision (i.e., 0 or 1). Therefore, we can calculate the probability of pairwise agreement within case $i$ as follows:

$$P_i = \frac{A_i}{N_i} \quad \text{—Eq. 2}$$

Suppose there are $m$ cases in total. Thus, the average probability of inter-reader agreement can be calculated as the mean of all $P_i$ as follows:

$$P_{avg} = \frac{1}{m}\sum_{i=1}^{m} P_i \quad \text{—Eq. 3}$$

To estimate uncertainty, we perform bootstrapping with $B$ replications. For each replication $b$, we randomly resample the list of $P_i$ values with replacement and compute the average agreement, $P_{avg}^{(b)}$. Our final estimate of the inter-reader agreement can then be calculated as the mean of all bootstrapped samples and the 95% confidence intervals (CI) are calculated using the percentile bootstrap approach:

$$\hat{P}_{avg} = \frac{1}{B}\sum_{b=1}^{B} P_{avg}^{(b)} \quad \text{—Eq. 4}$$

With a median of 18 radiologists reading each examination, a total of 64937 unique reader pairs were observed in this cohort. Setting $B$=1,000,000 and considering different splits of the cohort, we can estimate inter-reader agreement at the two different cut-offs as shown below in Table 1.

Table 1. Inter-reader agreement of radiologists estimated using 6174 readings from 62 radiologists (45 centers in 20 countries) across a retrospective cohort of 400 MRI examinations, considering different cut-offs and splits of the patient cohort.

| Threshold | Inter-reader agreement of radiologists (95% CI) | | |
| --- | --- | --- | --- |
| | All cases (n=400) | Negative cases (n=267) | Positive cases (n=133) |
| PI-RADS ≥3 | 73.4% (71.5%, 75.4%) | 65.8% (63.8%, 67.8%) | 88.7% (85.7%, 91.6%) |
| PI-RADS ≥4 | 78.2% (76.3%, 80.0%) | 73.8% (71.6%, 76.0%) | 86.9% (83.7%, 89.8%) |

We observe that the inter-reader agreement of radiologists is highly sensitive to the prevalence of disease in a given cohort. Thus, to account for prevalence shifts, we adjust our estimate of inter-reader agreement for each of the three trial testing cohorts as follows:

$$\hat{P}_{adj} = p \cdot \hat{P}_+ + (1-p) \cdot \hat{P}_- \quad \text{—Eq. 5}$$

Here, $\hat{P}_{adj}$ is the prevalence-adjusted general estimate of the inter-agreement of radiologists in a given testing cohort, $p$ is the prevalence of disease in this cohort, $\hat{P}_+$ is the general estimate for the inter-reader agreement of radiologists within positive cases, and $\hat{P}_-$ is the general estimate for the inter-reader agreement of radiologists within negative cases for a given cut-off.

- For the PRIME trial cohort, setting $p$=30% and considering the PI-RADS ≥3 cut-off, we estimate $\hat{P}_{adj}$ as **72.7%**.
- For the STHLM3-MRI trial cohort, setting $p$=17% and considering the PI-RADS ≥4 cut-off, we estimate $\hat{P}_{adj}$ as **76.0%**.
- For the IP1-PROSTAGRAM trial cohort, setting $p$=4% and considering the PI-RADS ≥4 cut-off, we estimate $\hat{P}_{adj}$ as **74.3%**.

## Estimation of Agreement between Radiologists and the Standard of Care Diagnoses

We estimate the probability that a randomly chosen radiologist diagnosing clinically significant prostate cancer on multiparametric MRI would agree to the standard of care diagnosis for a given case, using the outcomes of the PI-CAI multi-reader multi-case (MRMC) observer study. We consider the two same cut-offs of PI-RADS ≥3 and PI-RADS ≥4 (as used before to calculate inter-reader agreement) to threshold all radiology assessments into binary decisions of 0 or 1.

For a case $i$, let $n_i$ be the number of readers who scored the case and $A_i$ be the number of readers in agreement with the standard of care diagnosis. Therefore, we can calculate the probability of agreement between radiologists and the standard of care diagnosis within case $i$ as follows:

$$P_i = \frac{A_i}{n_i} \qquad \text{—Eq. 6}$$

Let the estimated probability that a randomly chosen radiologist agrees to the standard of care diagnosis within all cases be $\hat{Q}_{avg}$, within positive cases only be $\hat{Q}_+$, and within negative cases only be $\hat{Q}_-$. Following, Eq. 3–4 (as used before to calculate inter-reader agreement) setting $B$=1,000,000 and considering different splits of the cohort, we can derive general estimates for the agreement between radiologists and the standard of care diagnoses at the two different cut-offs as shown below in Table 2.

Table 2. Agreement of radiologists to the standard of care diagnoses estimated using 6174 readings from 62 radiologists (45 centers in 20 countries) across a retrospective cohort of 400 MRI examinations, considering different cut-offs and splits of the patient cohort.

| Threshold | Agreement of radiologists to the standard of care diagnoses (95% CI) | | |
| --- | --- | --- | --- |
| | All cases (n=400) | Negative cases (n=267) | Positive cases (n=133) |
| PI-RADS ≥3 | 68.6% (65.6%, 71.5%) | 58.1% (54.6%, 61.6%) | 89.5% (85.8%, 92.8%) |
| PI-RADS ≥4 | 76.5% (73.7%, 79.2%) | 72.4% (69.0%, 75.7%) | 84.6% (80.0%, 88.9%) |

Next, we apply Eq. 5 to derive prevalence-adjusted estimates ($\hat{Q}_{adj}$) for the probability that a randomly chosen radiologist agrees with the standard of care diagnosis within a given testing cohort.

- For the PRIME trial cohort, setting $p$=30% and considering the PI-RADS ≥3 cut-off, we estimate $\hat{Q}_{adj}$ as **67.5%**.
- For the STHLM3-MRI trial cohort, setting $p$=17% and considering the PI-RADS ≥4 cut-off, we estimate $\hat{Q}_{adj}$ as **74.6%**.
- For the IP1-PROSTAGRAM trial cohort, setting $p$=4% and considering the PI-RADS ≥4 cut-off, we estimate $\hat{Q}_{adj}$ as **73.0%**.

## Selection of AI Operating Point

We dichotomize all standard of care diagnoses into binary assessments of 0 (no evidence of clinically significant cancer on histopathology, if available, otherwise a PI-RADS score of 2 or lower from two expert urogenital radiologists in consensus) or 1 (Gleason grade group 2 or greater on histopathology) for evaluation. Similarly, we require an operating point to dichotomize all continuous 0–100 predictions from the AI system into binary assessments of 0 (negative for the likelihood of clinically significant cancer) or 1 (positive for the likelihood of clinically significant cancer), before we can evaluate its agreement with respect to the standard of care diagnoses.

### Matching Sensitivity

In the PI-CAI study, all AI-based predictions were thresholded to an operating point matching the sensitivity of the radiology readings at the PI-RADS ≥3 cut-off (89.4% sensitivity, 57.7% specificity) for evaluation. Our aim was to assess whether the AI system could reduce the number of false positives generated in comparison to a pool of radiologists, given that it was able to detect just as many clinically significant cancers as the radiologists. In this study, nearly all negative radiology readings (i.e., PI-RADS score of 2 or lower) across the external testing cohorts were exempt from verification (e.g., histopathology, follow-up outcomes) as per standard diagnostic practice. Due to imperfect reference standard bias present in our study, we anticipate that the operating point of radiology readings at the PI-RADS ≥ 3 cut-off within the external testing cohorts will exhibit a similar specificity (≈57.7%), but a biased estimate for sensitivity (≈100%).[32]

### Matching Specificity

Alternatively, by thresholding the AI system's predictions to match the ≈57.7% specificity of PI-RADS ≥3 readings, although we attain an operating point that demonstrates relatively more utility, such an operating point remains limited in comparison to the standard of care diagnoses due to its moderately low matching specificity. In Table 3 and Figure 1, we illustrate this using a simulated set of 1000 labels (representing the binary ground-truth labels of a randomly generated testing cohort with 30% disease prevalence), and five simulated sets of 1000 predictions (representing the 0–100 predictions of five separate AI systems that score an AUROC of 0.75, 0.80, 0.85, 0.90 or 0.99 on this testing cohort). When thresholding such a simulated AI system, even one that acts as a near perfect classifier (with an AUROC of 0.99), to an operating point matching the specificity of PI-RADS ≥3 readings, the maximum proportion of agreement that can be achieved with respect to the simulated standard of care diagnoses is limited to 69.9%.

### Youden's Index

When using the Youden's index as the operating point (i.e., a mathematically optimal cut-off that maximizes a balance of sensitivity and specificity), the proportion of agreement to the standard of care diagnoses scales all the way towards perfect agreement in the case of a perfect classifier. Therefore, in this study, we opt for using the Youden's index as the prespecified operating point for the AI system —solely for the purpose of assessing its overall diagnostic agreement with respect to the standard of care diagnoses. However, given that Youden's index is not a clinically meaningful operating point on its own and that there is no general "one size fits all" solution that is broadly applicable to all medical centers, our recommendation is to not use such a fixed operating point for clinical practice. Instead, we recommend that every medical center or adopter of such an AI system (that has been rigorously tested for its overall effectiveness), carefully considers their exact use-case (e.g., primary diagnosis, population-based screening), patient demographic (e.g., observed disease prevalence, distribution of disease stages) and downstream diagnostic capacity (e.g., available personnel, funding, number of sites, number of scanners), to calibrate and select a personalized operating point matching their specific needs.[33]

| AI System | Operating Point | AUROC | Sens | Spec | Agreement (Krippendorff's Alpha) | Proportion of agreement to the standard of care diagnoses |
|---|---|---|---|---|---|---|
| A | PR3$_{spec}$ | 0.75 | 0.77 | 0.57 | 0.24 | 63.2% |
| B | PR3$_{spec}$ | 0.80 | 0.83 | 0.57 | 0.28 | 64.9% |
| C | PR3$_{spec}$ | 0.85 | 0.90 | 0.57 | 0.33 | 67.0% |
| D | PR3$_{spec}$ | 0.90 | 0.95 | 0.57 | 0.36 | 68.5% |
| E | PR3$_{spec}$ | 0.95 | 0.98 | 0.57 | 0.38 | 69.5% |
| F | PR3$_{spec}$ | 0.99 | 1.00 | 0.57 | 0.39 | 69.9% |
| A | Youden | 0.75 | 0.64 | 0.73 | 0.33 | 70.0% |
| B | Youden | 0.80 | 0.68 | 0.76 | 0.40 | 73.2% |
| C | Youden | 0.85 | 0.75 | 0.78 | 0.49 | 77.2% |
| D | Youden | 0.90 | 0.82 | 0.83 | 0.61 | 82.5% |
| E | Youden | 0.95 | 0.86 | 0.89 | 0.73 | 88.2% |
| F | Youden | 0.99 | 0.95 | 0.95 | 0.88 | 95.1% |

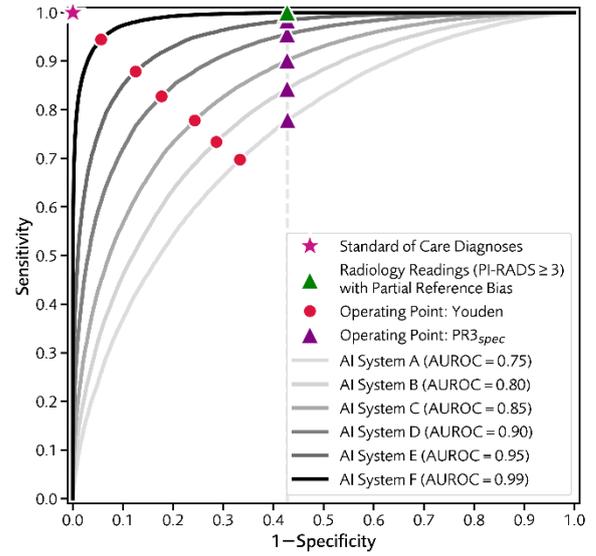

**Table 3.** (left), **Figure 1.** (right) Diagnostic sensitivity, specificity and agreement with respect to the standard of care diagnoses, for five simulated AI systems demonstrating different levels of overall performance (as measured by the AUROC metric), using a simulated testing cohort of 1000 cases with 30% prevalence. Each of the five models are thresholded at two different operating points: matching the 57.7% specificity of PI-RADS ≥3 readings (PR3$_{spec}$), and Youden's index (Youden).

## Interchangeability Criterion

We investigate the diagnostic interchangeability of the AI system to the radiology readings from routine practice with respect to their case-level predictions of Gleason grade group 2 or greater prostate cancer. To this end, we calculate the test statistic as the proportion of cases in each external testing cohort, where the binarized AI-based assessments are in agreement with the standard of care diagnoses of 0 (no evidence of clinically significant cancer on histopathology, if available, otherwise a PI-RADS score of 2 or lower on MRI from two expert urogenital radiologists in consensus) or 1 (Gleason grade group 2 or greater on histopathology). Two-sided 95% Wald confidence intervals are generated for this proportion (without adjustment for the variable used for stratification at randomization) using paired n-out-of-n bootstrapping with 1,000,000 replications, where the patient is used as the resampling unit. Interchangeability is concluded for each external testing cohort separately, if the lower boundary of the two-sided 95% confidence interval for the test statistic is greater than $\hat{Q}_{adj}$ (i.e., the estimate for the proportion of radiology readings that are in agreement with the standard of care diagnoses), adjusted by an absolute margin of 0.05. For example, considering that $\hat{Q}_{adj}$ for the PRIME trial cohort is 67.5%, the lower boundary of the test statistic must exceed 62.5% in order to conclude diagnostic interchangeability (as shown below in Fig. 2). To provide a holistic view of the evidence and its clinical context, we present $\hat{P}_{adj}$, i.e., the estimate for inter-reader agreement of radiologists, alongside these statistical outcomes.

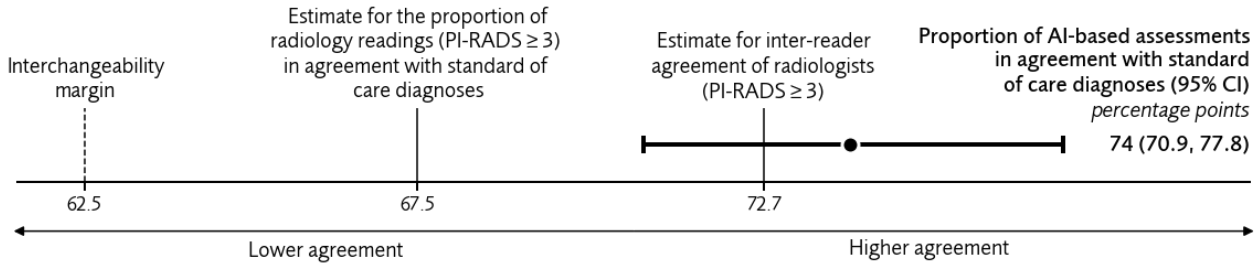

**Figure 2.** Simulation of the statistical analysis plan, considering the estimate for inter-reader agreement at the PI-RADS ≥3 cut-off as 72.7%, the estimate of radiology readings in agreement with the standard of care diagnoses at the PI-RADS ≥3 cut-off as 67.5%, an absolute interchangeability margin of 0.05, and a placeholder value for the observed proportion of AI-based assessments in agreement with the standard or care diagnoses of 74% (95% CI: 70.9%, 77.8%). Here, the lower boundary of the test statistic, i.e., 70.9%, exceeds the prespecified margin at 62.5%, thereby concluding diagnostic interchangeability.

## Secondary Measures

### Multiple Imputation Adjustment for Verification Bias

We primarily use the AUROC metric, adjusted for verification bias, to report all secondary measures. We adjust for the verification bias using multiple imputation.[32] Disease status (or the ground-truth label) will be imputed at the patient level for all negative MRI examinations lacking any histopathology evidence for verification, i.e., 24% (238 of 1000 cases) of the internal testing cohort, 33% (156 of 476 cases) of the PRIME external testing cohort, 67% (765 of 1143 cases) of the STHLM3-MRI external testing cohort, and 59% (229 of 391 cases) of the IP1-PROSTAGRAM external testing cohort. Based on the outcomes of the PI-CAI study, STHLM3-MRI trial (repeated screening round), and a recent meta-analysis of 42 studies using 7321 patients, we assume the base probability of clinically significant prostate cancer on a negative MRI examination as 0.03.[34,35] Using a logistic regression model trained on the outcomes of the fully-verified PI-CAI testing cohort (1000 multi-center, multi-vendor MRI examinations), we update this base probability considering the patient age and PSA level for each unverified examination within the testing cohorts of this study. For each candidate examination lacking verification, m=100 possible disease statuses are imputed, and in turn, an AUROC metric is computed m times using each set of imputations. Overall estimate of the AUROC metric will then be the mean of the m individual estimates of AUROC (denoted as $Q$). Total variance (denoted as T) will be calculated as the between-imputation variance (denoted as B) times (1+1/m) plus the within-imputation variance (i.e. mean of the m individual variances of the AUROC metric, denoted as $U$), as shown below [36]:

$$T = (B \times (1 + \frac{1}{m}) + U) \quad\quad\quad \text{—Eq. 7}$$

We compute 95% CI on the pooled AUROC metric as $Q \pm t_{m-1}\sqrt{T}$, where $\sqrt{T}$ is the pooled standard error and $t_{m-1}$ is sampled from the Student's t-distribution with m-1 degrees of freedom.

## Interim Analysis

As an interim analysis, we validate the PI-CAI-1 and PI-CAI-2B AI systems using the secondary internal testing cohort of 1000 MRI examinations (1000 patients; 22 cities in 13 countries). Performance for both AI systems will be reported using the AUROC metric (adjusted for verification bias) on a cohort-level basis and when stratified by center (or city). We hypothesize that despite being trained using a large-scale dataset (9,107 cases), the PI-CAI-1 system will demonstrate partial generalization or a drop in diagnostic performance when there is significant data drift, i.e., when the AI system is applied to a target population (multiethnic, intercontinental cohort, whose imaging were acquired using 34 scanner types from six MRI manufacturers) that deviates substantially from the distribution of its training dataset (homogenous population from four centers in two countries, whose imaging were acquired using nine scanner types from two MRI manufacturers). We hypothesize that the PI-CAI-2B system, which will be partially trained using non-overlapping patient data spanning all centers in the internal testing cohort, will demonstrate better overall performance in the absence of data drift. However, we anticipate that its diagnostic performance for certain centers may still remain limited due to external factors that influence the accuracy of prostate cancer detection (e.g., cohort composition, imaging quality).

As an additional measure of generalization and for further context in this analysis, we will evaluate the diagnostic performance of the PI-CAI-2B in comparison to the PI-CAI-1 system across the primary testing cohort of 1000 examinations (1000 patients; four cities in two countries) from the PI-CAI study.[20] Performance for both AI systems will be reported using the AUROC metric (without adjustment for verification bias) on a cohort-level basis and when stratified by center (or city).

## Subset Analyses

- **Subset analysis A:** We will measure the AUROC of the PI-CAI-2B system across subgroups of patients in the external testing cohorts, stratified by age as per the following categories: <50, 50–59, 60–69, and ≥70.

- **Subset analysis B:** We will measure the AUROC of the PI-CAI-2B system across subgroups of patients in the external testing cohorts, stratified by image quality using the Prostate Imaging–Quality version 2 (PI-QUAL v2) scoring system, as per the following categories: PI-QUAL 1 (inadequate), PI-QUAL 2 (acceptable), and PI-QUAL 3 (optimal).

- **Subset analysis C:** We will measure the AUROC of the PI-CAI-2B system across subgroups of patients in the external testing cohorts, stratified by patient ethnicity using a simplified representation of the OMB Directive 15 classification system, as per the following categories: White; Asian; Black, African or Caribbean; mixed or multiple.[26]

## Additional Analysis

In adherence with the expert recommendations of the PI-RADS steering committee, we will report the following metrics and benefit-to-harm ratios for all three external testing cohorts [37,38]:

- Sensitivity, specificity
- Positive predictive value, negative predictive value
- Concordance of AI-based assessments to the historical radiology readings and the standard of care diagnoses (represented via a 2×2 confusion matrix per comparison per external testing cohort)
- Gleason grade group ≥2 (number of true positives) to Gleason grade group 1 detections (number of false positives with Gleason grade group 1 diagnosis)
- Gleason grade group ≥2 (number of true positives) to Gleason grade group 1 (number of false positives with Gleason grade group 1 diagnosis) and no prostate cancer detections (number of negative predictions)
- Biopsy avoidance (number of true negatives) to Gleason grade group 1 detections (number of false positives with Gleason grade group 1 diagnosis)

Analysis will be conducted with respect to the PI-CAI-2B system's case-level predictions of Gleason grade group ≥2 cancer (thresholded according to Youden's index and at an operating point matching 90% sensitivity) and the ground-truth labels for a given external testing cohort (adjusted for verification bias).

## Exploratory Analyses

### Adjudication of False Positives and False Negatives

Further analysis will be conducted in conjunction with an expert radiologist (MdR; 12 years of experience in reading prostate MRI) at the central coordinating center, and members of the scientific advisory board (if deemed necessary), to adjudicate any systematic diagnostic errors made by the AI system across patient examinations potentially due to confounding factors (e.g., granulomatous prostatitis) or an inability to provide accurate diagnose across certain subtypes of cancer (e.g., mucinous adenocarcinoma) or under certain conditions (e.g., inadequate high b-value imaging).

### Sample Size and Statistical Power

As a measure of statistical power and to justify the adequacy of the sample size for each of the three external testing cohorts, we calculate the expected width of the 95% Wald CI of the primary endpoint, i.e., the proportion of AI-based assessments in agreement with the standard of care diagnoses. More specifically, for a given external testing cohort with N samples, we use the following formula to estimate the half-width of the 95% CI, $X$, for the primary endpoint, $p$, assuming a normal distribution with a critical value of 1.96 : [39]

$$X = 1.96 \times \sqrt{p \times \frac{1-p}{N}} \quad \text{—Eq. 8}$$

Assuming that the magnitude of $p$ is expected to fall between 70–80% (refer to Table 3), we can estimate $X$ as shown below:

**Table 4.** Expected half-width of the 95% confidence intervals (CI), $X$, for an expected magnitude of the primary endpoint, $p$ (i.e., the proportion of AI-based assessments in agreement with the standard of care diagnoses), given the sample size, N, for each of the three external testing cohorts (PRIME, STHLM3-MRI, IP1-PROSTAGRAM)

| External Testing Cohort | Expected half-width of the 95% CI, $X$, for an expected magnitude of the primary endpoint, $p$ | | |
|---|---|---|---|
| | $p$ = 70% | $p$ = 75% | $p$ = 80% |
| PRIME (N=476) | ±4.12% | ±3.89% | ±3.59% |
| STHLM3-MRI (N=1143) | ±2.67% | ±2.51% | ±2.32% |
| IP1-PROSTAGRAM (N=391) | ±4.54% | ±4.29% | ±3.96% |

Under these assumptions, the expected half-widths of the 95% CI are approximately ±3.59–4.12% for the PRIME external testing cohort (476 cases), ±2.32–2.67% for the STHLM3-MRI external testing cohort (1143 cases), and ±3.96–4.54% for the IP1-PROSTAGRAM cohort (391 cases), as seen in Table 4. These results indicate that, with the planned sample sizes, we will be able to estimate primary endpoint with acceptable certainty (i.e., < ±5%), supporting the statistical validity of our evaluation.

Anindo Saha [1,2], Joeran S. Bosma [1,3,4], Jasper J. Twilt [1,2], Alexander B.C.D. Ng [5,6], Aqua Asif [5,6], Kirti Magudia [7], Peder Larson [8], Qinglin Xie [9], Xiaodong Zhang [9], Chi Pham Minh [10], Samuel N. Gitau [11], Ivo G. Schoots [4,12], Martijn F. Boomsma [13], Renato Cuocolo [14], Nikolaos Papanikolaou [15,16], Daniele Regge [17,18], Derya Yakar [4,19], Mattijs Elschot [20,21], Jeroen Veltman [22,23], Baris Turkbey [24], Nancy A. Obuchowski [25], Jurgen J. Fütterer [2], Anwar R. Padhani [26], Hashim U. Ahmed [27,28], Tobias Nordström [29,30], Martin Eklund [29], Veeru Kasivisvanathan [5,6,31], Maarten de Rooij [32], Henkjan Huisman [1,20], on behalf of the PI-CAI–ProCAncer-I–COMFORT–STHLM3-MRI–PRIME consortia*

[1] Diagnostic Image Analysis Group, Radboud University Medical Center, Nijmegen, The Netherlands
[2] Minimally Invasive Image-Guided Intervention Center, Radboud University Medical Center, Nijmegen, The Netherlands
[3] Department of Health and Information Technology, Ziekenhuisgroep Twente, Almelo, The Netherlands
[4] Department of Radiology, Netherlands Cancer Institute, Amsterdam, The Netherlands
[5] Division of Surgery and Interventional Science, University College London, London, UK
[6] Centre for Urology Imaging, Prostate, AI and Surgical Studies Research Group, University College London, London, UK
[7] Abdominal Imaging Division, Department of Radiology, Duke University, Durham, NC, USA
[8] Department of Radiology and Biomedical Imaging, University of California, San Francisco, CA, USA
[9] Department of Medical Imaging, The Third Affiliated Hospital, Southern Medical University, Guangzhou, China
[10] Department of Diagnostic Radiology, Diagnostic Imaging Center, 108 Military Central Hospital, Hanoi, Vietnam
[11] Department of Radiology, Aga Khan University Hospital, Nairobi, Kenya
[12] Department of Radiology and Nuclear Medicine, Erasmus University Medical Center, Rotterdam, The Netherlands
[13] Department of Radiology, Isala Hospital, Zwolle, The Netherlands
[14] Department of Medicine, Surgery and Dentistry, University of Salerno, Baronissi, Italy

[15] Computational Clinical Imaging Group, Centre for the Unknown, Champalimaud Foundation, Lisbon, Portugal
[16] Department of Radiology, Royal Marsden Hospital, Sutton, United Kingdom
[17] Department of Radiology, Candiolo Cancer Institute, FPO-IRCCS, Candiolo, Turin, Italy
[18] Department of Surgical Sciences, University of Turin, Turin, Italy
[19] Department of Radiology, University Medical Center Groningen, Groningen, The Netherlands
[20] Department of Circulation and Medical Imaging, Norwegian University of Science and Technology, Trondheim, Norway
[21] Department of Radiology and Nuclear Medicine, St. Olavs hospital, Trondheim University Hospital, Trondheim, Norway
[22] Department of Radiology, Ziekenhuisgroep Twente, Hengelo, The Netherlands
[23] Department of Multi-Modality Medical Imaging, Technical Medical Centre, University of Twente, Enschede, The Netherlands
[24] Molecular Imaging Branch, National Cancer Institute, Bethesda MD, USA
[25] Department of Quantitative Health Sciences and Department of Diagnostic Radiology, Cleveland Clinic Foundation, Cleveland OH, USA
[26] Paul Strickland Scanner Centre, Mount Vernon Cancer Centre, London, UK
[27] Imperial Prostate, Division of Surgery, Department of Surgery and Cancer, Imperial College London, London, UK
[28] Imperial Urology, Division of Cancer, Cardiovascular Medicine and Surgery, Imperial College Healthcare NHS Trust, London, UK
[29] Department of Medical Epidemiology and Biostatistics, Karolinska Institutet, Stockholm, Sweden
[30] Department of Clinical Sciences, Danderyd Hospital, Karolinska Institutet, Stockholm, Sweden
[31] Department of Urology, University College London Hospitals NHS Foundation Trust, London, UK
[32] Department of Medical Imaging, Radboud University Medical Center, Nijmegen, The Netherlands

*A list of all consortia members and their affiliations will be published in an appendix alongside the study outcomes.